\documentclass{amsart}
\usepackage{graphicx}
\usepackage{multirow}

\usepackage{color}

\theoremstyle{definition}

\theoremstyle{remark}

\numberwithin{equation}{section}
\usepackage{subfig} 

\begin{document}

\title[Physicist's approach to public transportation networks]{Physicist's approach to public transportation networks: between data processing and statistical physics}


\author{Yaryna Korduba}
\address{Ukrainian Catholic University, Lviv 79011, Ukraine;
$\mathbb{L}^4$  Collaboration \& Doctoral College for the Statistical
Physics of Complex Systems, Leipzig-Lorraine-Lviv-Coventry, Europe}
\curraddr{}
\email{korduba\textunderscore y@ucu.edu.ua}
\thanks{Paper submitted to the Festschrift devoted to Prof.Dr. Jurij Kozicki on the occasion
of his 70th birthday}

\author{Yurij Holovatch}
\address{Institute for Condensed Matter Physics, National Acad. Sci. of Ukraine, Lviv 79011, Ukraine;
Coventry University, Coventry, CV1 5FB, United Kingdom;
$\mathbb{L}^4$  Collaboration \& Doctoral College for the Statistical
Physics of Complex Systems, Leipzig-Lorraine-Lviv-Coventry, Europe}
\curraddr{}
\email{hol@icmp.lviv.ua}
\thanks{}

\author{Robin de Regt}
\address{Coventry University, Coventry, CV1 5FB, United Kingdom;
$\mathbb{L}^4$  Collaboration \& Doctoral College for the Statistical
Physics of Complex Systems, Leipzig-Lorraine-Lviv-Coventry, Europe}
\curraddr{}
\email{deregtr@uni.coventry.ac.uk}
\thanks{}



\date{}

\dedicatory{}

\begin{abstract}
 
In this paper we aim to demonstrate how physical perspective enriches usual statistical
analysis when dealing with a complex system of many interacting agents of non-physical origin. 
To this end, we discuss analysis of urban public transportation networks viewed as 
complex systems. In such studies, a multi-disciplinary approach is applied by integrating 
methods in both data processing  and statistical physics to investigate the correlation 
between public transportation network topological features and their operational stability. 
The studies incorporate  concepts of coarse graining and clusterization, universality 
and scaling, stability and percolation behavior, diffusion and fractal analysis.

\end{abstract}

\maketitle

\section{Introduction}\label{I}

One of the most prominent scientists of our time, Steven Hawking, when asked about the 
main trends of science in the XXI century replied that he thinks it will be the century of complexity (see \cite{Sengupta06}). Indeed, 
it is becoming more apparent how complexity has gradually become one of the central concepts of modern science and, on a more general scale, of the whole human culture \cite{Ladyman13,Thurner16}. Complex systems of many interacting agents share an essential common property: 
they display collective behavior that does not follow trivially from the behaviors of their individual parts. Moreover, their behavior 
is characterized by a set of inherent features that include self-organization, emergence of new functionalities, extreme sensitivity to 
small variations in their initial conditions, and governing power laws (fat-tail behavior, see \cite{Holovatch17}). 
Complex system science -- a new emerging field --  aims to understand such behavior from a unified perspective and to formulate its description in a quantitative and predictable manner.

The methodological and conceptual framework of complex system science originates from many traditional disciplines, statistical physics being probably one of the most important ingredients \cite{Sherrington10,Anderson72,Parisi99,Goldenfeld99}. Another important ingredient is a universal language that allows for the description of various complex systems. This language originates from graph theory and is currently widely known as complex network science, see e.g. \cite{Albert02,Dorogovtsev03,Newman06,Russo17} and references therein. The latter serves as a framework to formalize a system of interacting agents by allowing each agent to act as network node, where various kinds of interaction can be described as (weighted or unweighted, directed or undirected, multiple or unique) links. The former equips complexity science with the whole arsenal of tools and concepts traditionally used in physics to describe collective phenomena.

In the case study discussed in this paper, we aim to demonstrate how a physical perspective can enrich statistical analysis and data processing when dealing 
with a complex system of many interacting agents `non-physical' in nature. To this end, we have chosen to consider public transportation networks where the 
data for each public transport network (PTN) includes information about the routes and the locations of each station. Using 
this information, the question is, what new insights can one gain by appealing to physical concepts and ideas for its analysis?

The set-up of the paper is as follows: in the next section \ref{II} a short review of some of the papers where a public transportation system have 
been analyzed from a complex system science perspective is provided. Then concepts of coarse graining and clusterization of PTNs are further discussed 
in section \ref{III}. In the subsequent sections we discuss the search for universal features in network structure and the power laws (scaling) 
governing node-degree distributions (section \ref{IV}); analysis of a networks behavior to failure of their parts and its resemblance to percolation phenomenon 
(section \ref{V}); the phenomenon of diffusion and its contribution to PTN modeling and studies of fractal properties of transportation networks (section \ref{VI}). 
Finally, in section \ref{VII} we conclude with some final statements and outlook.

It is our pleasure and honor to submit this paper to the Festschrift dedicated to Prof.Dr. Jurij Kozicki on the occasion of his 70th birthday. His contributions to the analysis of collective behavior in condensed matter, to complex network and complex system science are well known and highly appreciated. This in fact motivated us in choosing the topic for this paper.


\section{PTN from complex network perspective: a review}\label{II}

Analysis of PTNs as complex systems relies on their presentation in the form of a graph, i.e. of a complex network. Such analysis began comparatively recently. One of the first papers appeared in 2002, where the topological properties of the Boston subway \cite{latora2002boston} were analysed. Subsequently, similar analysis has been performed for many other PTNs around the world. The types of PTNs investigated include the subway \cite{latora2002boston,Seaton2004},
bus \cite{xu2007scaling,sui2012space,yang2011bus,guo2013scaling}, rail \cite{sen2003small}, air 
\cite{guimera2004modeling,guimera2005worldwide,guida2007topology} and various combinations of these 
\cite{Ferber09a,Sienkiewicz05,soh2010weighted,alessandretti2016user}.

\begin{figure}[h]
\begin{center}
\includegraphics[width=12cm]{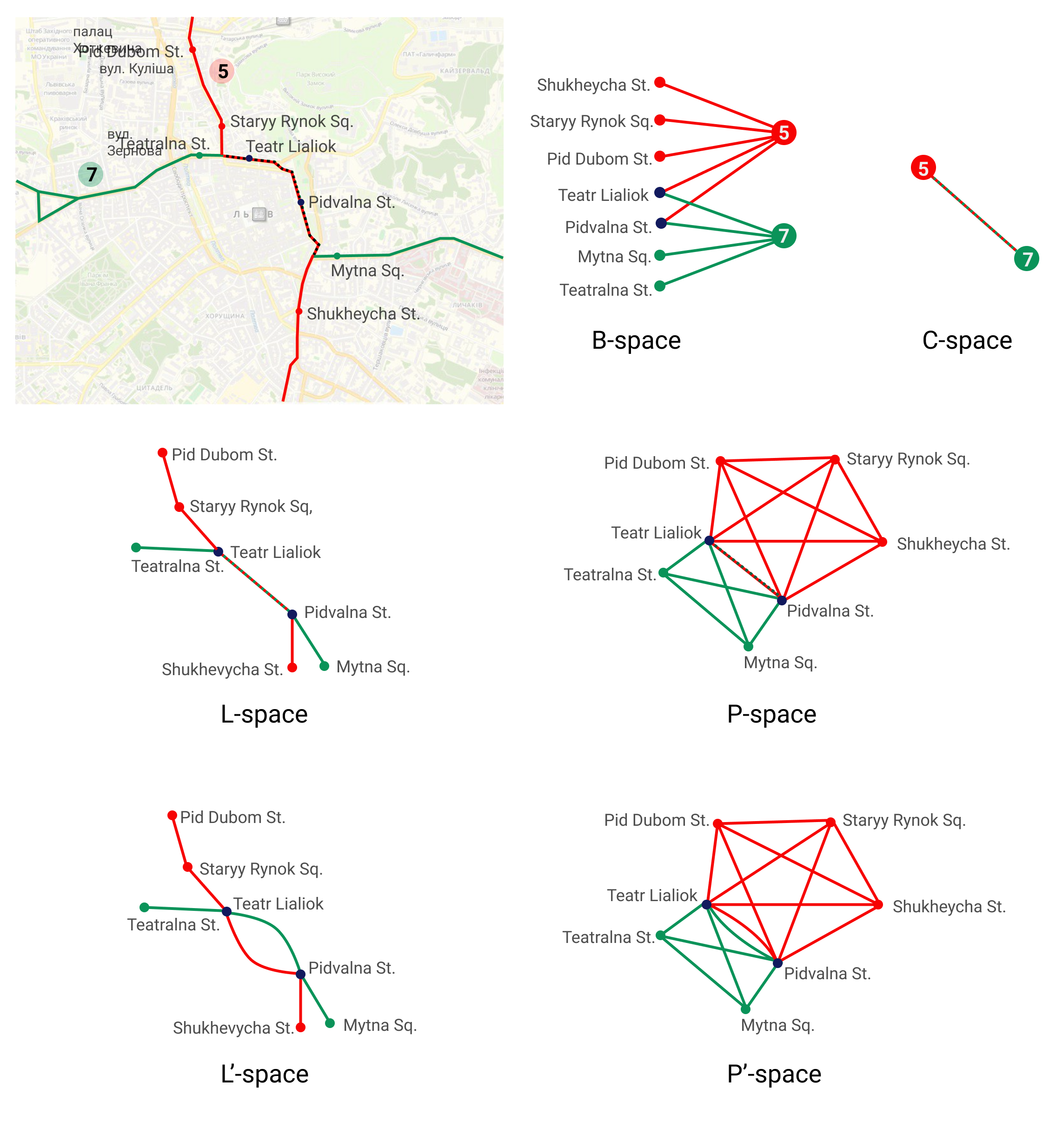}
\caption{A subsection of the Lviv PTN with two routes (Nos 5 and 7) and seven stops and its different graph representations, $\mathbb{L}$-, $\mathbb{L'}$-,
$\mathbb{P}$-, $\mathbb{P'}$-, $\mathbb{B}$-, and $\mathbb{C}$-`spaces'. See text for more
discussion.} \label{fig2.1}
\end{center}
\end{figure}

The general goal of these and other similar studies was to present a PTN in a form of a graph (complex network) and studying different features of the graph presentation to gain more information about the properties of a PTN. Thus far, a number of different topological representations of a PTN have been developed, by attributing different constituents of a PTN to vertices (interchangeable with nodes) and edges (interchangeable with links) of the corresponding graph. To give an example, in Fig. \ref{fig2.1} 
we show a subset of a PTN map of Lviv (Ukraine) and some of the graphical representations. In the $\mathbb{L}$-space representation, the stations are 
represented by graph nodes, the nodes are connected if the corresponding stations are adjacent in a route whereas multiple links are substituted by a singe one. 
The corresponding $\mathbb{L'}$-space representation keeps multiple links. In particular, it allows one to study the so-called `harness' behavior of PTN routes. This concept describes how different routes tend to follow similar paths for a certain number of stations. The harness distribution $P(r,s)$ can be defined as the number of sequences of consecutive stations $s$, serviced by $r$ parallel routes \cite{Ferber09a,berche2009network}. A similar feature has also been studied for weighted networks in Ref.\cite{xu2007scaling}. 

The $\mathbb{L}$-space topology is ideal for studying the connectivity of networks which is crucial for PTN operation. To this end, network metrics such as mean shortest 
path length or largest connected component size serve as import indicators of PTN operation. This space has been applied in many different studies of real world networks \cite{Ferber09a,Sienkiewicz05,latora2002boston}.

In the $\mathbb{P}$-space representation, similar to $\mathbb{L}$-space, each station is presented as a node however here links join together all nodes that belong to a particular route and form a complete subgraph. Different subgraphs are then joined via common 
stations, that share different routes. This representation has been applied in many studies 
\cite{sen2003small,Seaton2004,Sienkiewicz05,Ferber09a,xu2007scaling,ghosh2010structure}. It is useful, in particular for determining the mean number of vehicle changes when traveling between any two points on a PTN service network. Similar to the 
$\mathbb{L}$-space topology, in $\mathbb{P}$-space multiple links do not exist. 

In the so-called $\mathbb{B}$-space \cite{Ferber09a,chang2007assortativity} one constructs a bipartite graph that contains nodes of two types: node-stations and node-routes. Naturally, the single-mode projection of the $\mathbb{B}$-space graph to the nodes-stations leads to $\mathbb{P}$-space. In turn, an analogous projection to the nodes-routes leads to the so-called $\mathbb{C}$-space \cite{Ferber09a}. Here, one considers how routes are connected to each other and describes how routes are linked throughout the network. In $\mathbb{C}$-space if any two routes service the same station they are obviously linked.

These and similar presentations of PTNs in complex network form enable one to answer different questions about the functional features of PTNs of many cities and to study the relationship of these features with the topology of corresponding complex networks. The topics of analysis included understanding the collective phenomena taking place on the network, in particular analysis of PTN robustness to random break down or targeted removal of their constituents 
\cite{berche2009resilience,berche2012transportation,vonFerber12}, 
development of a number of simulated growth models for PTNs \cite{berche2009network,yang2011bus,sui2012space},
studies of their spatial embedding \cite{sui2012space,Benguigui91,benguigui1995fractal,benguigui1992fractal,vonFerber13,Ferber09a}.
This list is far from being complete \cite{Holovatch11a,deRegt17}. In general, such analysis has revealed that PTNs constructed in cities with different geographical, cultural and historical background share a number of basic common topological properties. They appear to be strongly correlated structures with high values of clustering coefficients and comparatively low mean shortest path values. Their node degree distributions are often found to follow exponential or power-law decay. 
Moreover, some of these observables can be employed as  key performance indicators in aid of further developing efficient and stable PTNs. Referring the interested reader to the original publications, in the forthcoming sections we plan to use some of the previously obtained results together with the new data (mainly for PTNs of Lviv and Bristol \cite{Yarynka}) to show how the physical perspective enriches analysis of complex PTN networks.

\section{Coarse graining in systems of interacting particles vs PTN
clusterisation}\label{III}

In statistical physics, coarse graining reduces the number of degrees of freedom in a system of interacting particles thus enabling its analytical treatment or computer 
simulation. In a coarse grained description, one smooths over fine structure of a system passing to a new system with `rescaled' constituents and interactions between them. Sometimes, the properties of such rescaling enable one to describe the large-scale behavior of a system as a whole,
see e.g. \cite{Kadanoff66,Stanley71,Kozitsky88,Holovatch_books}. A similar technique is also at hand as an effective method to simplify complex network descriptions
in public transportation analysis \cite{gallotti2015multilayer,Yarynka}. 

\begin{table}[ht]
\begin{center}
\setlength{\tabcolsep}{4pt}
\begin{tabular}{l r r r r r r r}
 \hline
  City & Population  & Area  & $N$ & $R$  &  Vehicle  \\
   &   & km$^2$ & &  &  types \\
   \hline
    Lviv & 721 301 & 182 & 768 &   77 & BET   \\
    Bristol & 535 907 & 110 & 1474  & 143   &  BF \\
  \hline
\end{tabular} 
\end{center}
\caption{General information about the cities and their PTNs. $N$: number of stops, 
$R$: number of routes,  vehicle types:  B (bus), E (electric trolley), T (tram), 
F (ferry). \label{tab1}
}
\end{table}

Here we provide an example of how coarse graining is applied in preparing a (simplified) database for PTN analysis. As a case study, we demonstrate this taking the PTNs 
of Lviv (Ukraine) and Bristol (UK). Both cities and their PTNs are of comparable size, see Table \ref{tab1}, and we use some of the results obtained for their PTNs as 
illustrations in this and forthcoming sections \cite{deRegt17,deRegt19,Yarynka}. 

One of the common features of urban PTNs is the existence of stations that are situated in close proximity to one another. Thus, very often there is no need to consider such stops separately. One can use this fact to simplify the transport networks before analysis. To prepare the database, one can merge the stops that fall within a reasonable walking distance. To this end, a density-based clustering algorithm DBSCAN \cite{ester1996density} has been applied to coarse grain nodes in Lviv and Bristol PTNs in the  $\mathbb{L}$-space representation \cite{Yarynka}. 
This algorithm considers a minimal clustering radius $R$ and the minimal number of the points 
to organize a cluster. The algorithm divides all the points in a network into different categories:\footnote{Core points together with their density reachable and directly density reachable points form the clusters. Noise points in the particular 
case of stops clusterization were not rejected.} 
\begin{itemize}
    \item Core points. A point $p$ is called a core point if it lays within a clustering radius $R$ from a core point $c$,
    \item Density reachable points.
    A point $p$ is called reachable from a core point $c$ if there exists a chain of points $p_1, p_2, ..., p$, $p_1=c$ and a 
    point $p_{i-1}$ is directly reachable from $p_i$,
    \item Noise. A point $p$ is called a noise point if it does not belong to any cluster.
\end{itemize}


\begin{figure}[h!]
\begin{center}
\includegraphics[height=6cm]{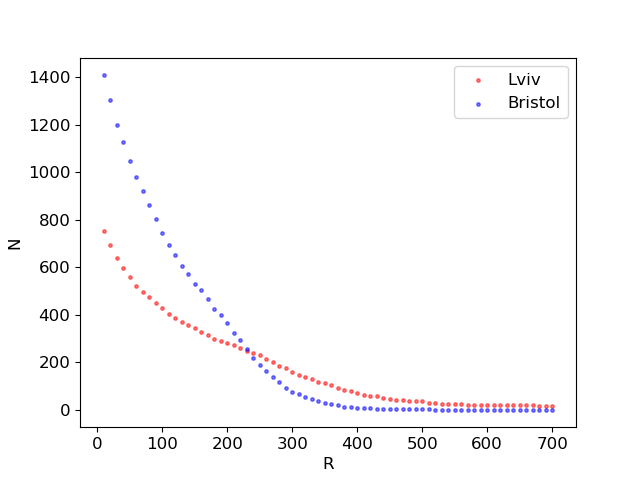}  
\caption{Coarse graining of network nodes in Lviv and Bristol PTNs: dependence of the stops number $N$ 
on the clustering radius $R$ (in meters). \label{fig3.1}} 
\end{center}
\end{figure}

\begin{figure}[ht!]
\begin{center}
\subfloat[R=30m]{
\includegraphics[width=6cm]{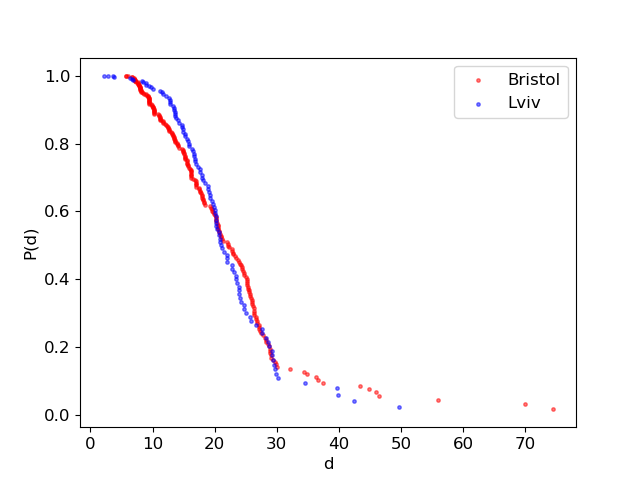}  
}
\subfloat[R=40m]{\includegraphics[width=6cm]{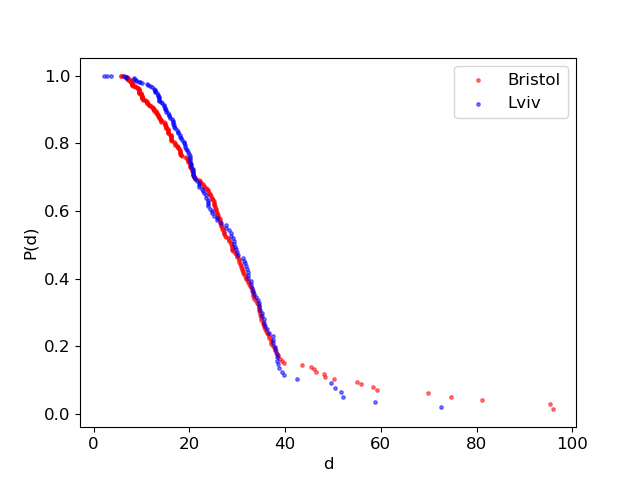}   
}

\subfloat[R=50m]{\includegraphics[width=6cm]{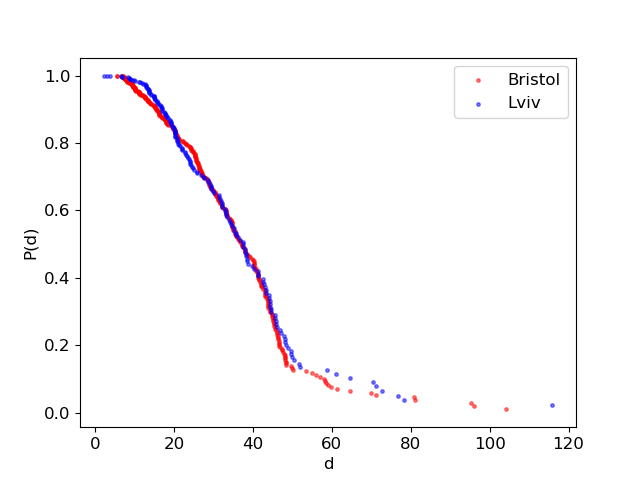}   
}
\subfloat[R=60m]{\includegraphics[width=6cm]{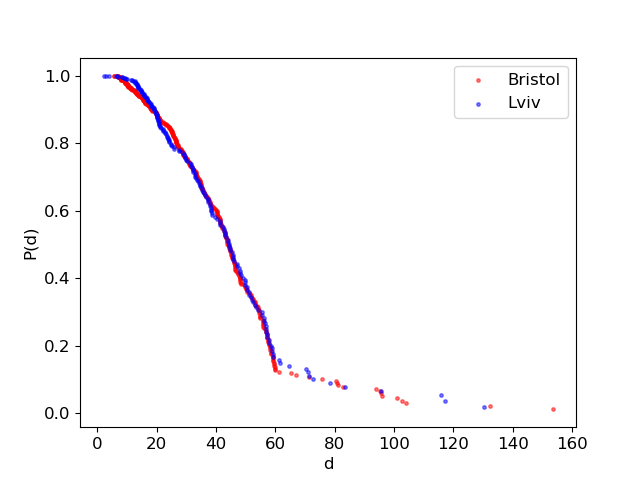}  
}
\caption[Cumulative distribution of maximal distances between the stops in each cluster]
{Cumulative distribution $P(d)$ of maximal distances  $d$ between stops of each cluster in the PTNs of Lviv and Bristol for four different values
of clustering radious $R$.\label{fig3.2}}
\end{center}
\end{figure}

To receive meaningful results of the analysis, one should properly choose the clustering radius $R$. If $R$ is too small, no stops will be grouped into clusters and if $R$ is too large, all the stops will form one cluster, see Fig. \ref{fig3.1}. Moreover, the maximal distance between the stops in a cluster should not exceed a reasonable walking distance. Checking maximal inter-node distance $d$ in a cluster provides additional criterion to choose $R$. Indeed, with $d$ too high one gets unrealistic situations when stops that are further away from 
each other are considered as a single node. In Fig. \ref{fig3.2} we show the cumulative distribution of maximal distances $d$ between the stops of each cluster for different values of $R$. As one can see from Fig. \ref{fig3.2}, at $R=40$ m the maximal distances in clusters do not exceed a reasonable walking distance of 100 m. This suggests such value of $R$ to be optimal and used in the coarse-graining of two networks under discussion.

Note that the PTN coarse-graining procedure discussed in this section heavily relies of the (Euclidean) distance between network nodes. This is a typical feature of complex networks embedded in space. Besides PTNs, other examples of spatial networks are given by other transportation networks, power grids, neural networks and much more \cite{Barthelemy11}. For spatial networks, analysis of the topology alone can not disclose their properties
and is to be completed by corresponding analysis in geographical space. We will provide more insight on this issue in Section \ref{VI}.

\section{Universality and scaling}\label{IV}

The prevailing sections of this paper involve explicitly or implicitly the concept of universality. The search for universal - i.e. independent on system details - 
features of wide classes of systems is inherent to physics. In statistical physics, universality means that typical behavior of systems consisting of many interacting 
parts is independent of system's structural details. In complex system science, such analysis very often means the search for common statistical laws that govern the 
behavior of systems of many interacting agents. It has been found that in many cases such laws attain power law asymptotes: probability of a rare event decays slower 
than predicted by a central limit theorem. Special systematization is being suggested, 
it classifies power-law statistics observed in different systems according to certain archetypal reasons that cause 
it \cite{Mitzenmacher03,Newman05,Simkin11,Corominas-Murtra15}. In this respect it is similar to distinguishing different universality classes in the theory of critical 
phenomena, see e.g. \cite{Kozitsky04,Holovatch_books}. Such power laws have been observed for the PTNs too, as we show in several examples given below. The main reasons that cause their appearance in PTN statistics are due to the so called preferential attachment and optimization scenarios. The optimization scenario was suggested by Benoit Mandelbrot \cite{Mandelbrot53} based on information theory. Within this scenario, power laws appear as a result of optimization of the information transmitted and its costs. Within the preferential attachment, in the course of system evolution new elements tend to create links with those, which already have more links. Sometimes such a scenario is named rich get richer. One of the most known models where such a scenario is realized was suggested
by Herbert Simon \cite{Simon55}. 

As it was mentioned in Section \ref{II}, there are different ways to present a PTN in a complex network form. Correspondingly, the nodes and links of the complex network have different interpretations in different formalisms, 
different `spaces' described in Section \ref{II}. An important quantity that characterizes a PTN in all these representations is the node degree distribution $p(k)$. 
The function $p(k)$ gives the probability that an arbitrary chosen node of the network has $k$ links. For the classical Erd\"os-R\'enyi random graph of finite size $N$, $p(k)$ is given by the binomial distribution \cite{Bollobas01}. It tends to a Poisson distribution as $N\to \infty$. The last rapidly decays for large $k$ and is characterized by a typical scale. A similar property is shared by other distributions that are characterized by an exponential decay:
\begin{equation}\label{4.1}
p(k)\sim e^{-k/\hat{k}}, \, k\to \infty
\end{equation}
-- they are characterized by a typical scale, a decay constant $\hat{k}$. Such a property does not hold if the distribution possesses power-law asymptotics:
\begin{equation}\label{4.2}
p(k)\sim k^{-\gamma}, \, k\to \infty \, .
\end{equation}
Function (\ref{4.2}) is a particular case of a more general class of homogeneous functions \cite{Stanley71} that share the following property:
\begin{equation}\label{4.3}
p(ak) = a^{\lambda} p(k) \, ,
\end{equation}
for all nonzero $a$. Exponent $\lambda$ is called the homogeneity degree. Obviously, $\lambda=-\gamma$ for $p(k)$ given by (\ref{4.2}). 

\begin{figure}[ht!]
	\centerline{
		\subfloat[]{
			\includegraphics[width=6cm,angle=90]{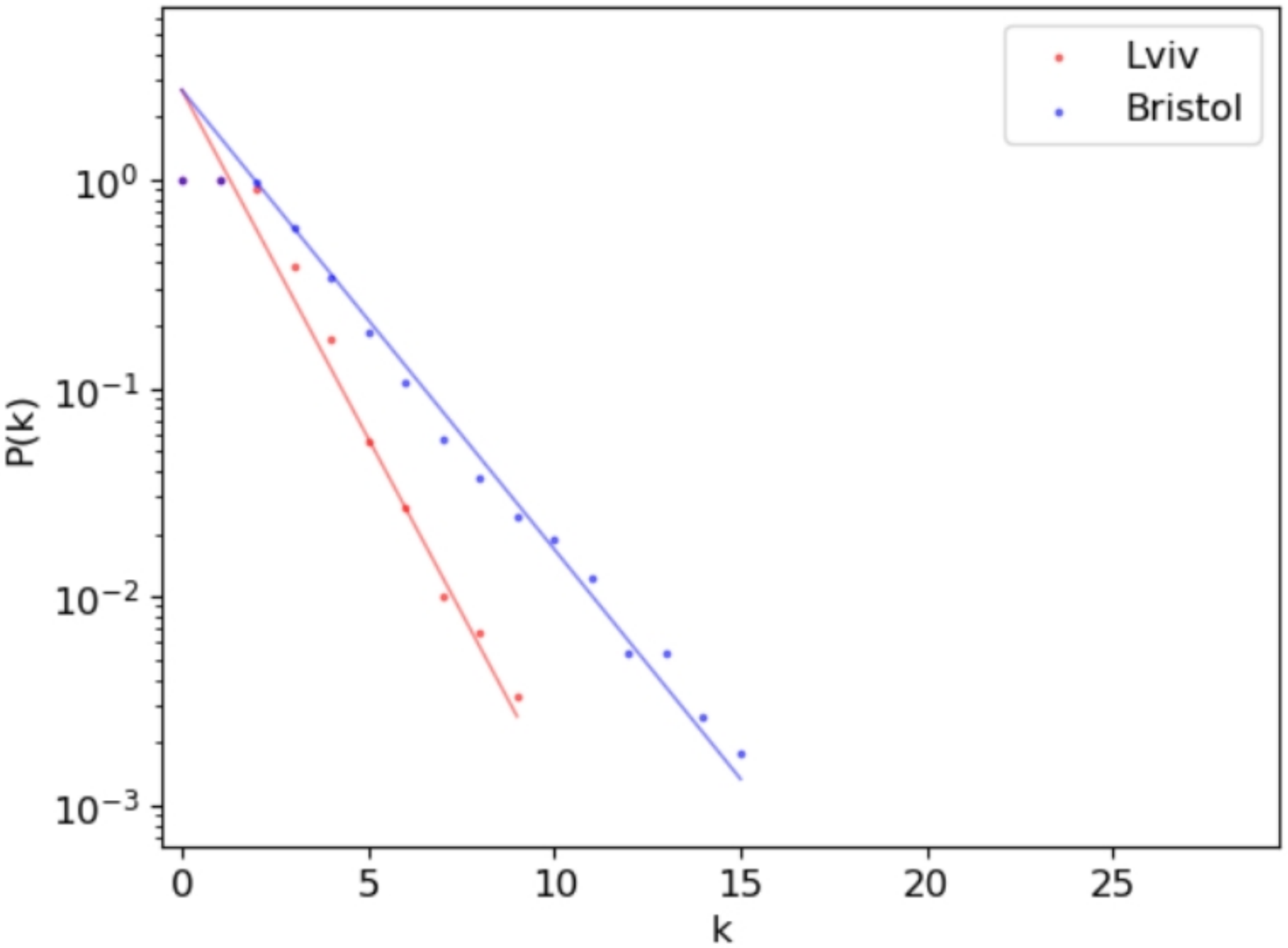}  }
		\subfloat[]{\includegraphics[width=6cm,angle=90]{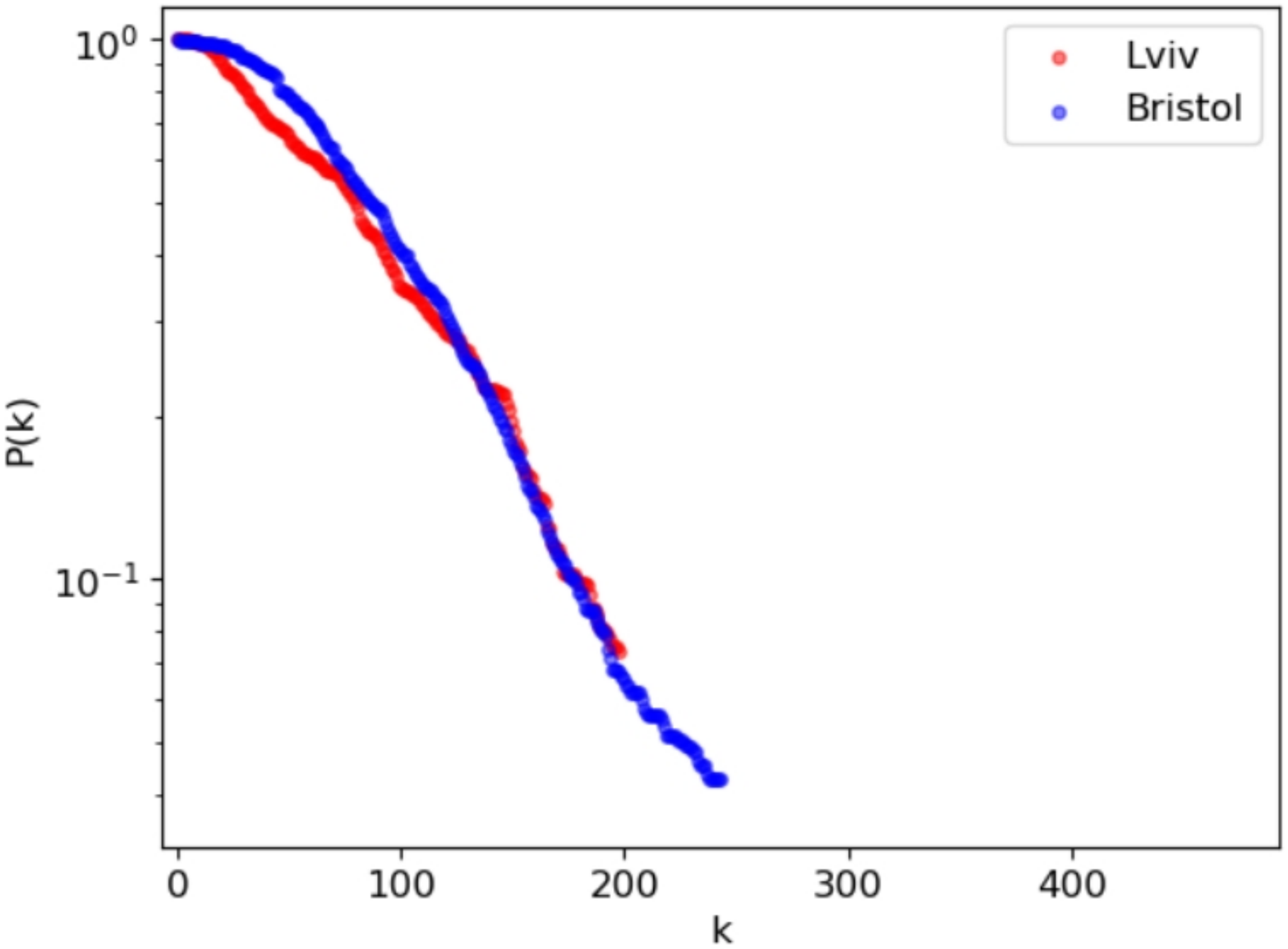} }
		}
				\caption[Cumulative node degree distributions for Lviv and Bristol PTN]
		{Cumulative node degree distributions $P(k)$ for Lviv and Bristol PTN. A: $\mathbb{L}$-space, B: $\mathbb{P}$-space \cite{Yarynka}.\label{fig4.1}}
	
\end{figure}

\begin{figure}[ht!]
	\centerline{
		\subfloat[]{
			\includegraphics[width=6cm]{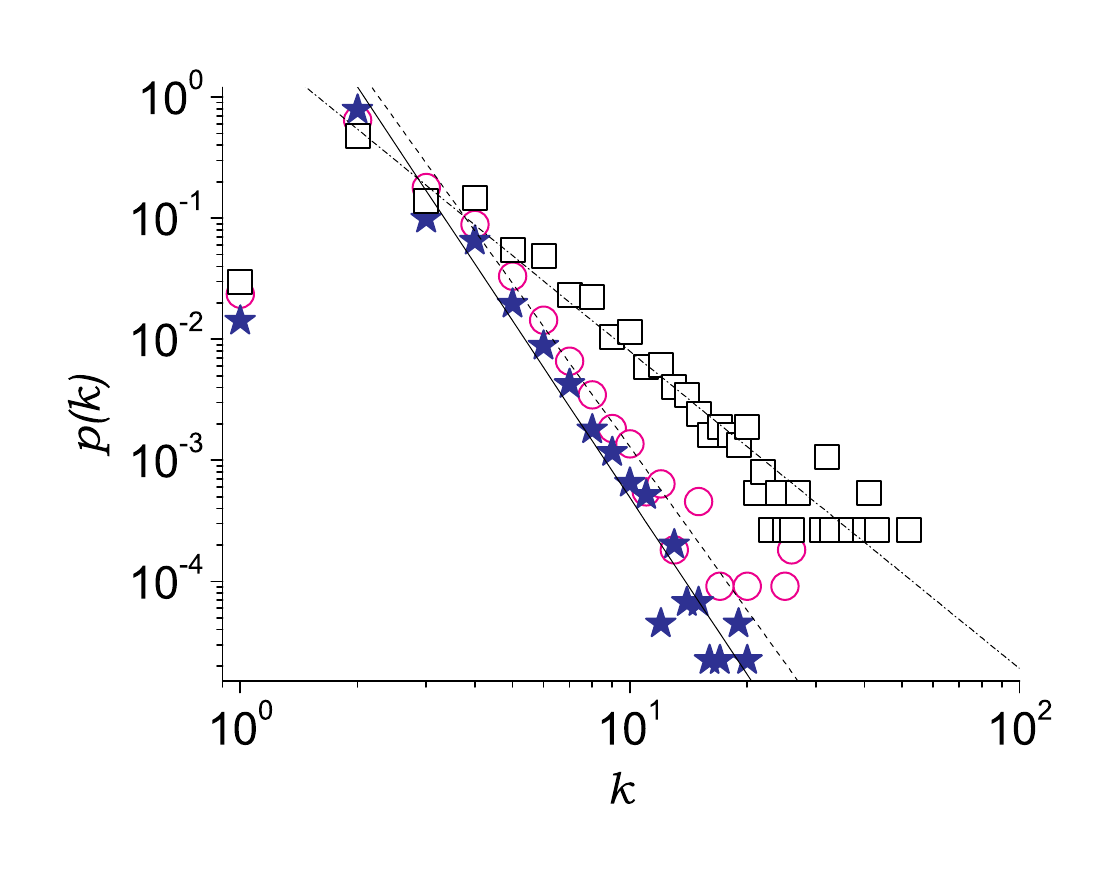}  }
		\subfloat[]{\includegraphics[width=6cm]{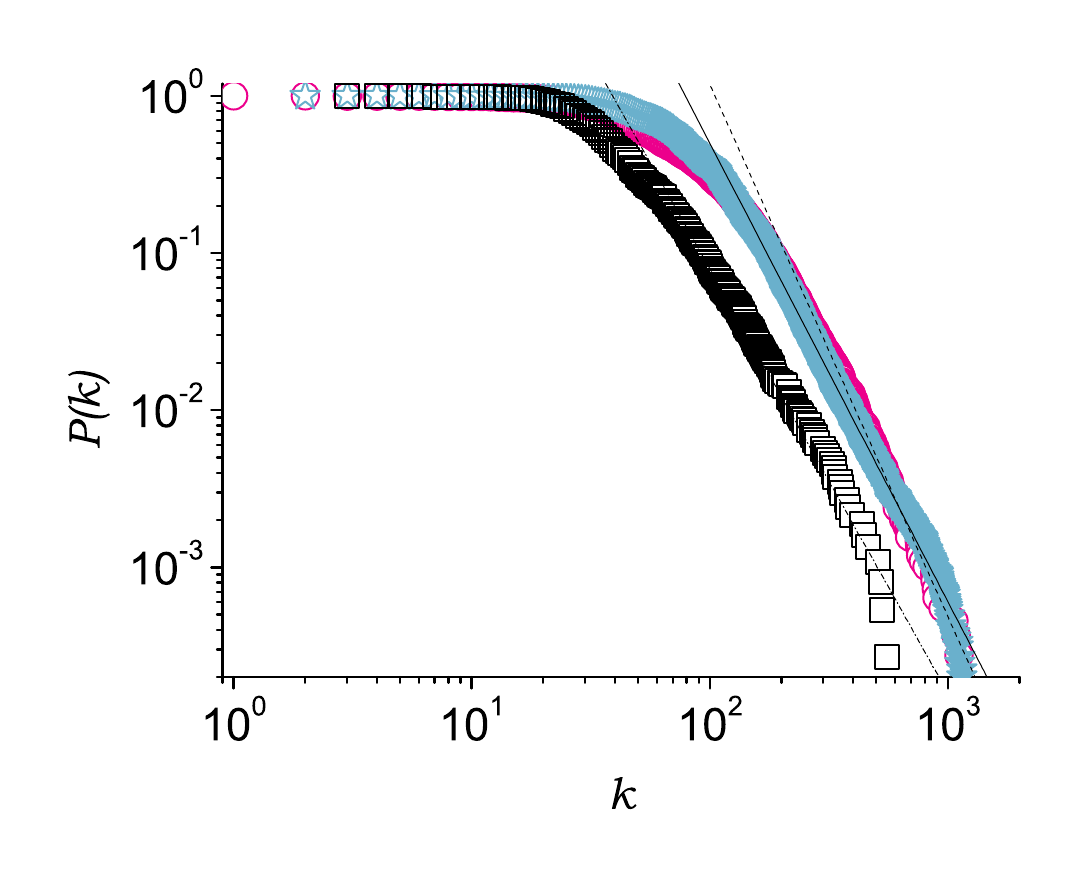} }
	}
	\caption[Node degree distributions $p(k)$ for London, Los Angeles and Paris PTN]
	{(A): Node degree distributions of the PTNs of London (circles),	
	Los Angeles (stars) and Paris (squares) in the $\mathbb{L}$-space. (B): cumulative node degree distributions 
	of the same cities in the $\mathbb{P}$-space  \cite{Ferber09a}.\label{fig4.2}}
	
\end{figure}

Networks that are characterized by $p(k)$ satisfying (\ref{4.3}) are called scale-free. Examples of scale-free behavior have been found in many networks that 
represent natural and man-made systems, see e.g. \cite{Albert02,Dorogovtsev03,Newman06,Russo17} and references therein. There are some signals of scale-free behavior 
in PTNs too \cite{von2007network,Ferber09a}. In Fig. \ref{fig4.1} we show cumulative node degree 
distributions $P(k)=\sum_{q=k}^{k_{\rm max}}p(q)$ for Lviv and Bristol PTNs \cite{Yarynka}. As it can be seen from the figures, both plots can be fitted reasonably
well in a log-linear plot via a linear dependency: corresponding $P(k)$ functions manifest exponential decay. However, for some PTNs, power laws hold.
Due to obvious spatial constraints power-law behavior is observed in the $\mathbb{L}$-space for rather low values of $k$. In Fig. \ref{fig4.2} it is demonstrated for PTNs of several 
cities, with the exponents  $\gamma_{\mathbb{L}}=4.48$ (London), $\gamma_{\mathbb{L}}=4.85$ (Los Angeles) and  $\gamma_{\mathbb{L}}=2.62$ (Paris) \cite{Ferber09a}. 
In $\mathbb{P}$-space however, the construction of a network allows for much higher node degrees as far as each route enters the network as a complete graph of 
constituting stations. Cumulative node-degree distributions for the same cities manifest scale-free behavior in a large region of $k$ with $\gamma_{\mathbb{P}}=3.89$
(London),
$\gamma_{\mathbb{P}}=3.92$ (Los Angeles), and  $\gamma_{\mathbb{P}}=3.70$ (Paris)
\cite{Ferber09a}. A similar effect may be reached by coarse graining in $\mathbb{L}$-space:
joining a cluster of stops to a single station one naturally increases the degree of the coarse grained station. In turn, this may lead to the scale-free behavior \cite{Shanmukhappa16}.

Another peculiar feature of scale-free networks that makes their properties so different
from the networks with an exponentially decaying node degree distribution is the behavior of their distribution moments. 
For exponentially decaying $p(k)$, all moments $\langle k^m \rangle$ are finite even in the limit $N\to \infty$.
However, this is not the case for scale-free networks. Indeed for an infinite network with $p(k)$ given by (\ref{4.2}) only lower
moments $\langle k^m \rangle$ with $m<\lambda-1$ converge. As we will see in the forthcoming sections, this feature leads to essential 
effects even in the case of a finite-size PTN.

\section{Percolation vs PTN stability}\label{V}

Percolation is an archetypal example of collective behavior. It finds analogies in different many-particle systems besides phenomenon of fluid percolating in a porous media and outside physics in its classical sense too \cite{Essam80,Stauffer91}. Different phenomena occurring on complex networks are related to percolation. What methods are available which allow for the construction of a network containing a giant cluster and moreover what would be the properties of the network when this cluster appears? For a given connected network, what are the strategies to destroy a giant cluster, and how robust is the
network when different attack scenarios are applied? These and similar questions give
rise to the application of percolation theory and its ideas in the network description and answering them sheds light on many common features shared between 
percolation and phenomena that occur on complex networks \cite{berche2009resilience,Holovatch06,Kozitsky14}. 
 
Percolation as a concept has been widely used in PTNs analysis to answer questions about their stability to different types of events \cite{berche2009resilience,von2009attack,berche2012transportation,vonFerber12}, see also 
Refs. \cite{Holovatch11a,deRegt17} for a review. Such events influence the operational properties and consequently decrease the connectivity of a network. In critical cases, they lead to overall network collapse. One discriminates between two categories of harmful events in a network: random failures and targeted attacks. Random failures might be caused by car accidents, weather conditions, substantial traffic jams, etc. Targeted attacks include terrorist acts, strikes, etc. Targeted attacks usually occur at the most important parts of the network to cause the most critical damages. Attack simulation provides an effective way for network stability assessment. A common process to model different types of attacks on PTN is to remove constituents of the corresponding graph according to certain rules (attack scenario) and to study network robustness to such removals. Different indicators are at hand to monitor network robustness. The inverse mean shortest path length and the size of the largest connected components are the most common ones.

\begin{figure}[]
\begin{center}
\subfloat[Bristol, $\mathbb{L}$-space]{
\includegraphics[width=6.5cm]{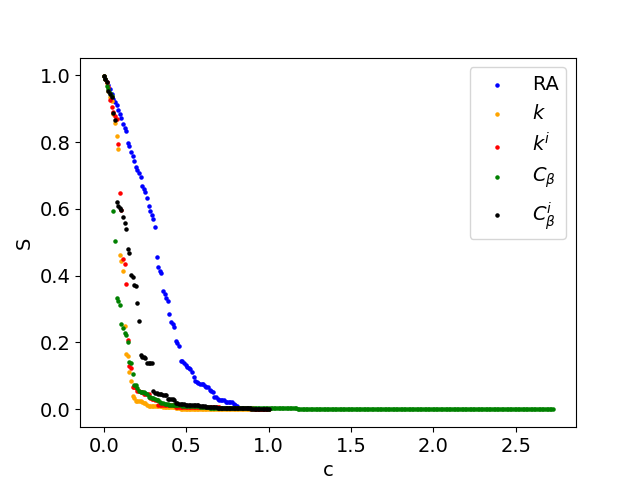} \label{attackSimulations:f1}
}
\subfloat[Lviv, $\mathbb{L}$-space]{\includegraphics[width=6.5cm]{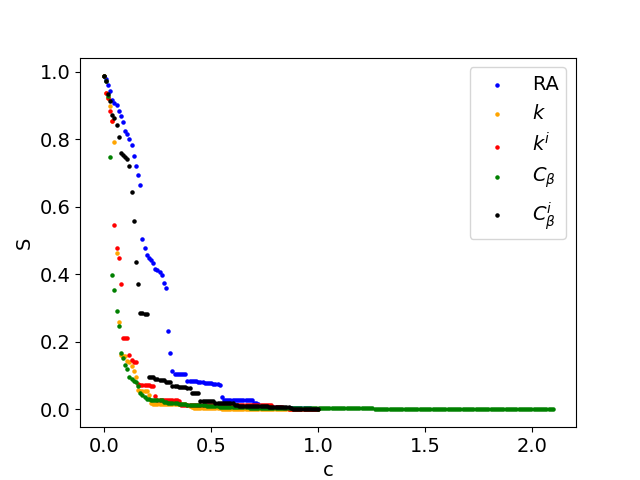}   \label{attackSimulations:f2}
}

\subfloat[Bristol, $\mathbb{P}$-space]{\includegraphics[width=6.5cm]{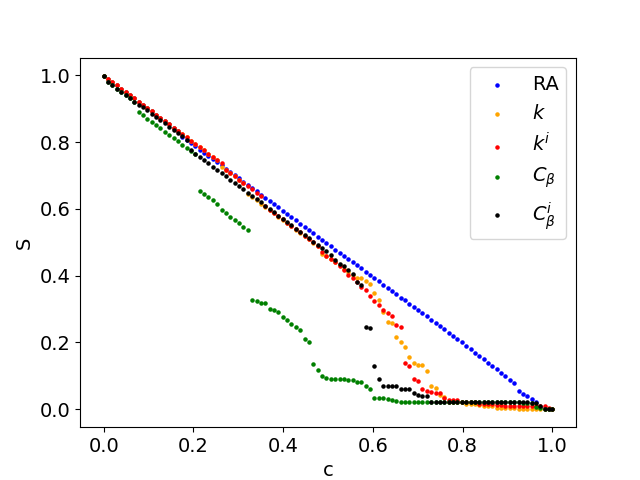}   \label{attackSimulations:f3}
}
\subfloat[Lviv, $\mathbb{P}$-space]{\includegraphics[width=6.5cm]{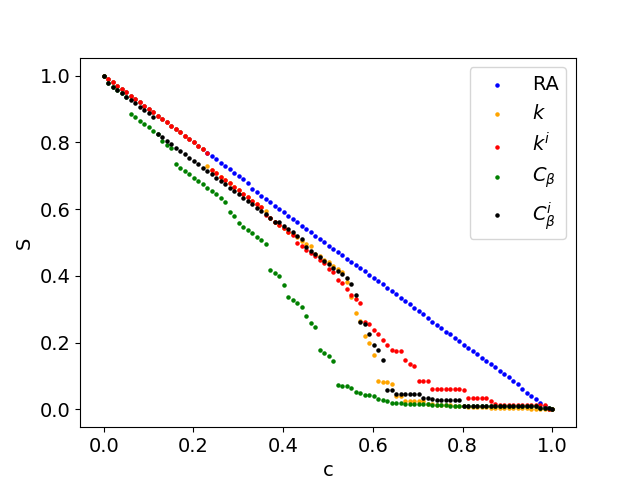}   \label{attackSimulations:f4}
}

\subfloat[Bristol, $\mathbb{P}$-space]{\includegraphics[width=6.5cm]{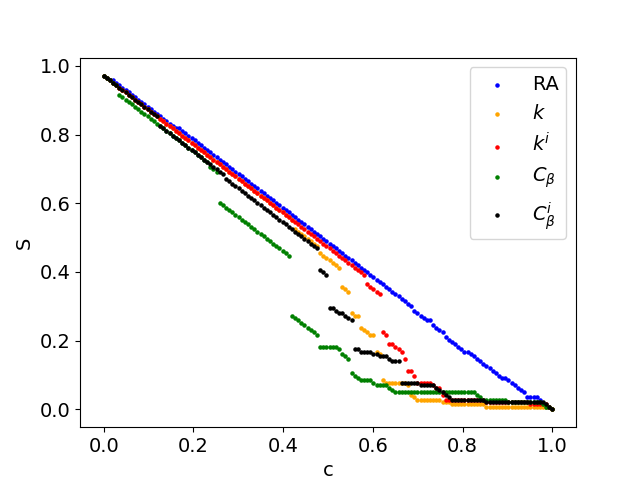}   \label{attackSimulations:f5}
}
\subfloat[Lviv, $\mathbb{P}$-space]{\includegraphics[width=6.5cm]{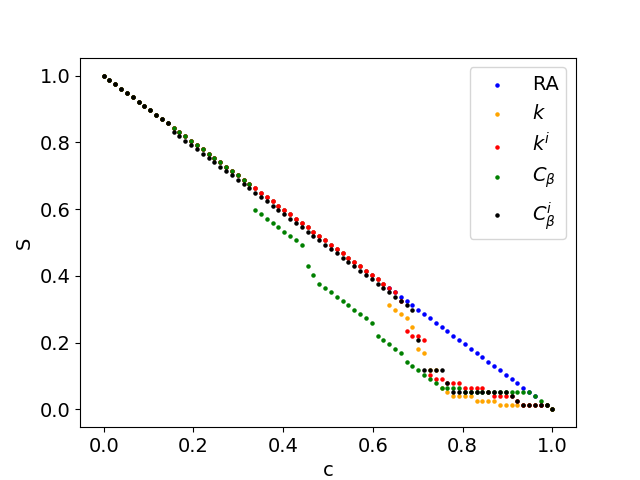}   \label{attackSimulations:f6}
}

\end{center}
\caption[Attacks on PTNs in $\mathbb{L}$- and $\mathbb{P}$-space]
{Changes in the largest connected component $S$ of Lviv and Bristol PTNs caused by removal (attack) of a share $c$ of the corresponding
complex network nodes in $\mathbb{L}$-, $\mathbb{P}$- and $\mathbb{C}$-spaces. Attacks are performed randomly ($RA$) or according to the node
lists ordered by the initial node degree ($k^i$), recalculated after each attack step node degree ($k$), 
initial betweenness centrality ($C_{\beta}^i$) and recalculated after each attack step betweenness centrality ($C_{\beta}$). \label{fig5.1}}
\end{figure}

As an example, in Fig. \ref{fig5.1} we demonstrate changes in the normalized size of the largest connected component $S$ of Lviv and Bristol PTNs caused by removing a share $c$ of their nodes. Due to network inhomogeneity such removal can be done in different ways. Within the simulation, during the random attack scenario (denoted {\em RA} in the plots) the nodes are removed at random. 
For the targeted attacks, the most important nodes are removed first. To this end, we have chosen two indices to evaluate node importance: the node degree $k$ and the node betweenness centrality $C_{\beta}$. The first index measures the number of links of a given nodes, the second measures the number of shortest paths between all other nodes of the network that go through the given one. The structure of the network changes during an attack, so the node indices do. Therefore, the nodes were removed either according to the lists prepared before the attack, or the lists were updated during each simulation step. Corresponding curves are denoted in the plots as $k$ and $k^i$ for the highest node degree attack scenario and $C_\beta$ and $C_\beta^i$ for the highest betweenness centrality scenario.

Plots of Fig. \ref{fig5.1} demonstrate typical features of behavior of complex networks under attacks: they are robust with respect to random removal of their constituents (slow decay of the curves at $RA$ scenario) and vulnerable to the targeted attacks. A useful
indicator of network robustness has been proposed in general context in Ref. \cite{Schneider11} and further exploited for PTN analysis in \cite{berche2012transportation}. To this end, one estimates the area $A$ under each $S(c)$ plot:
\begin{equation}
    \label{5.1}
    A = \int_{0}^{1} S(c) dc
\end{equation}
and further uses it to quantify network robustness to attacks: the larger the area, the more robust the network is. Corresponding values of $A$ for different attacks are given in Table \ref{tab5.1}.

\begin{table}[ht!]
\begin{center}
\setlength{\tabcolsep}{4pt}
\begin{tabular}{c | l r r r r r}
\hline
  Space & City & $RA$ & $k^{i}$ & $k$ & $C_{\beta}^{i}$ & $C_{\beta}$ \\
   \hline
    \multirow{2}{4em}{$\mathbb{L}$} & Bristol &0.304 &0.125 &0.109 & 0.159& 0.095\\
    & Lviv & 0.234 & 0.087 &0.075 &0.159 & 0.059  \\
    \hline
    \multirow{2}{4em}{$\mathbb{P}$} & Bristol & 0.498 & 0.438 & 0.439 & 0.416 & 0.31 \\
    & Lviv & 0.497& 0.423& 0.403&0.4 &0.321 \\
    \hline
    \multirow{2}{4em}{$\mathbb{C}$} & Bristol & 0.481 & 0.432 & 0.404 & 0.395 & 0.343 \\
    & Lviv & 0.498 & 0.47 & 0.464 & 0.465 & 0.426  \\
  \hline
\end{tabular} 
\end{center}
\caption[Lviv and Bristol PTNs robustness to attacks]
{Lviv and Bristol PTNs robustness  $A$ (\ref{5.1}) to attacks of different types. See  caption of Fig. \ref{fig5.1} for the explanation
of attack types. \label{tab5.1}}
\end{table}

The simulation results can give many useful insights. For example, the attack simulations for Lviv and Bristol PTNs showed that both cities in $\mathbb{L}$-, $\mathbb{P}$- and $\mathbb{C}$-space react in the same way to different types of attacks (see Fig. \ref{fig5.1}). The least harmful was the random attack simulation, while the most dangerous were attacks at the nodes with the highest betweenness centrality 
(with resorting of the `importance lists'). The simulations also showed that Bristol PTN is more resilient in $\mathbb{L}$-space than Lviv PTN. The last observation can be also explained on the base the analytic results available for infinite uncorrelated networks \cite{berche2009resilience}. The Molloy-Reed criterion \cite{molloy1995critical} allows to predict the stability of a network to random attacks having only a couple of its properties. It states, that the giant connected component is present in any uncorrelated network if 
\begin{equation}
\label{5.2}
\kappa = \frac{\langle k^2 \rangle}{\langle k \rangle} \geq 2,
\end{equation}
where $\langle k \rangle$ and $\langle k^2 \rangle$ stand for mean and mean square node degree correspondingly. The higher $\kappa$, the more stable the network. Although this criterion has been derived for infinite uncorrelated networks, 
the value of the Molloy-Reed parameter $\kappa$ has been used to predict the robustness of a PTN: one expects a network with a higher value of $\kappa$ to be more robust with respect to random removal of its parts
\cite{berche2009resilience,berche2012transportation}. Such an estimate is of most use to
evaluate network stability in $\mathbb{L}$-space. In $\mathbb{P}$- and $\mathbb{C}$-space, the networks are well-connected by the nature of construction, therefore the corresponding values of $\kappa$ are very high. In Table \ref{tab5.2} we show values of the Molloy-Reed parameter $\kappa$ (\ref{5.2}) for Lviv and Bristol PTNs. As one can see comparing Tables \ref{tab5.1} and \ref{tab5.2}, higher robustness of Bristol PTN in  $\mathbb{L}$-space correlates with the higher value of  $\kappa_{\mathbb{L}}$ as compared with the 
corresponding data for Lviv PTN.

\begin{table}[ht]
\begin{center}
\setlength{\tabcolsep}{4pt}
\begin{tabular}{l r r r r r}
 \hline
  City & $\kappa_{\mathbb{L}}$ & $\kappa_{\mathbb{P}}$ & $\kappa_{\mathbb{C}}$ \\
   \hline
    Bristol & 4.493 & 145.456 & 43.104 \\
    Lviv & 3.099 & 138.413 & 44.245    \\
  \hline
\end{tabular} 
\end{center}
\caption{Molloy-Reed parameter $\kappa$ (\ref{5.2}) for Lviv and Bristol PTNs in $\mathbb{L}$-, $\mathbb{P}$- and $\mathbb{C}$-spaces. \label{tab5.2}
}
\end{table}

\section{Diffusion, random walks, fractals and PTN modeling}\label{VI}

One more analogy with physical phenomenon that has been exploited within PTN analysis and
modeling is given by diffusion. Indeed, being generally understood as the spreading
of any (material or non-material) objects, diffusion theory is widely applied in different fields. Whereas in its original formulation 
diffusion is driven by the gradient of concentration, 
numerous applications of diffusion concept consider it in terms of random walks (RWs) \cite{Kozitsky18}. In such a formulation, the mean square distance $\langle {\mathcal R}^2 \rangle$ between the beginning and the end of a RW of $N$ steps scales as:
\begin{equation}\label{6.1}
 \langle {\mathcal R}^2 \rangle = N \, ,
\end{equation}
here each RW step is assumed to be of equal unit length. 
\begin{figure}[ht!]
	\centerline{
		\includegraphics[width=9cm]{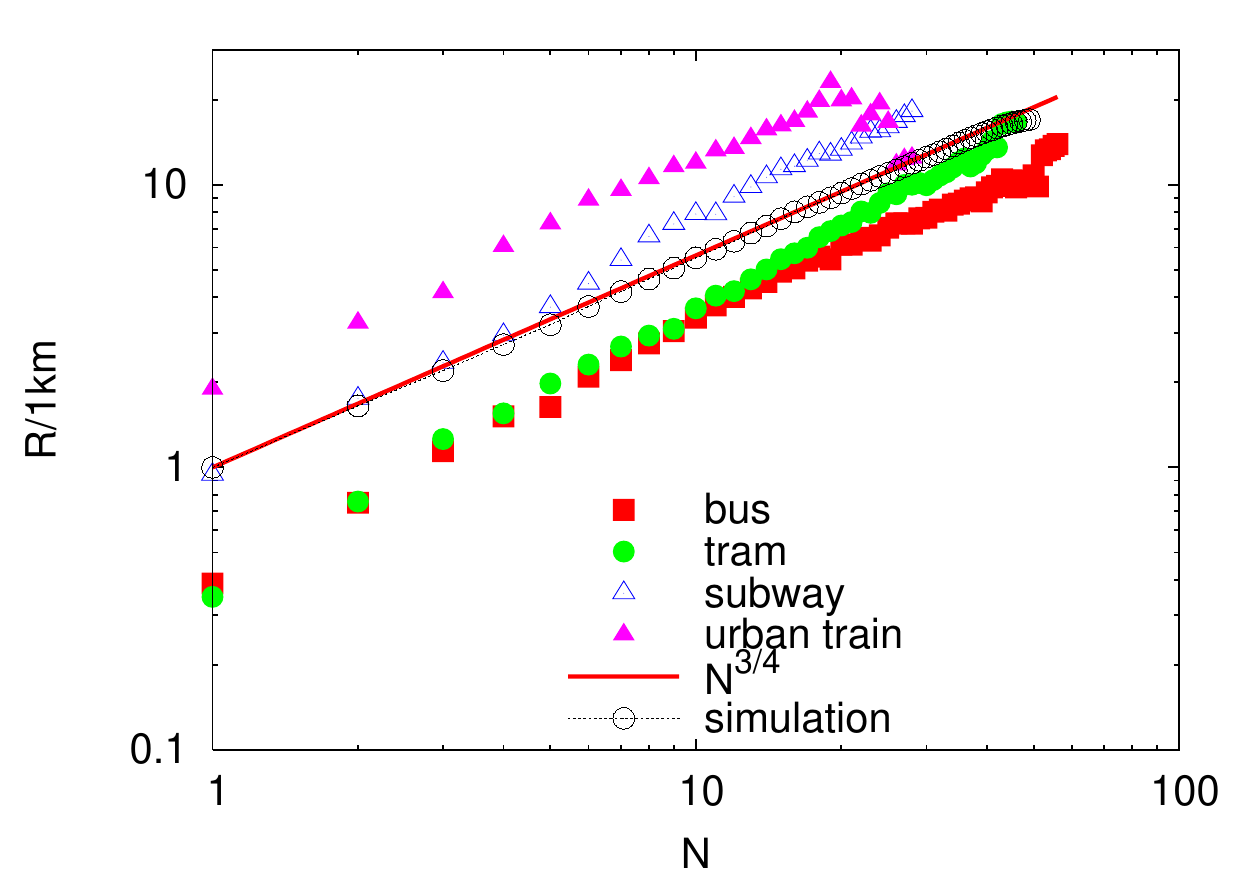} 
	}
	\caption[${\mathcal R}$ as function of the number of stations
	traveled]
	{Berlin PTN: Mean distance ${\mathcal R}$ as function of the number of stations
		traveled by different modes of transport in Berlin. Solid line shows result
		for a SAW at $d=2$ \cite{Nienhuis82} and for a simulated city \cite{von2007network}, see the text for more explanation. \label{fig6.1}}
\end{figure}

If one forbids a random walker to cross its trajectory (the so-call self-avoiding walk, SAW),
the scaling dependence still holds asymptotically, however with a different power law:
\begin{equation}\label{6.2}
\langle {\mathcal R}^2 \rangle \sim N^{2\nu} \, .
\end{equation}
Exponent $\nu$ is universal: it is the same for all SAWs on different lattices with the same 
dimensionality of space $d$. Obviously, $\nu=1/2$ for the RW in Eq. (\ref{6.1}). {For a SAW in $d=2$, the exact value of the exponent is known to be: $\nu=3/4$} \cite{Nienhuis82}. It is remarkable that similar scaling has been found for the mean square distance between different stations belonging to the same PTN route, see Fig. \ref{fig6.1} where this is demonstrated for different modes of transport in Berlin \cite{von2007network}. Later it was shown that the behavior of the route chains in a city is better described by the L\'evy flight process \cite{vonFerber13} rather than a SAW. Such empirical observation about SAW scaling of PTN routes seems to be counter-intuitive. One of possible explanation is that the shapes of such routes may result from an optimization with respect to total passenger traveling time and area coverage \cite{von2009modeling,Ferber09a}. 

\begin{figure}[ht!]
	\centerline{
		\includegraphics[width=9cm]{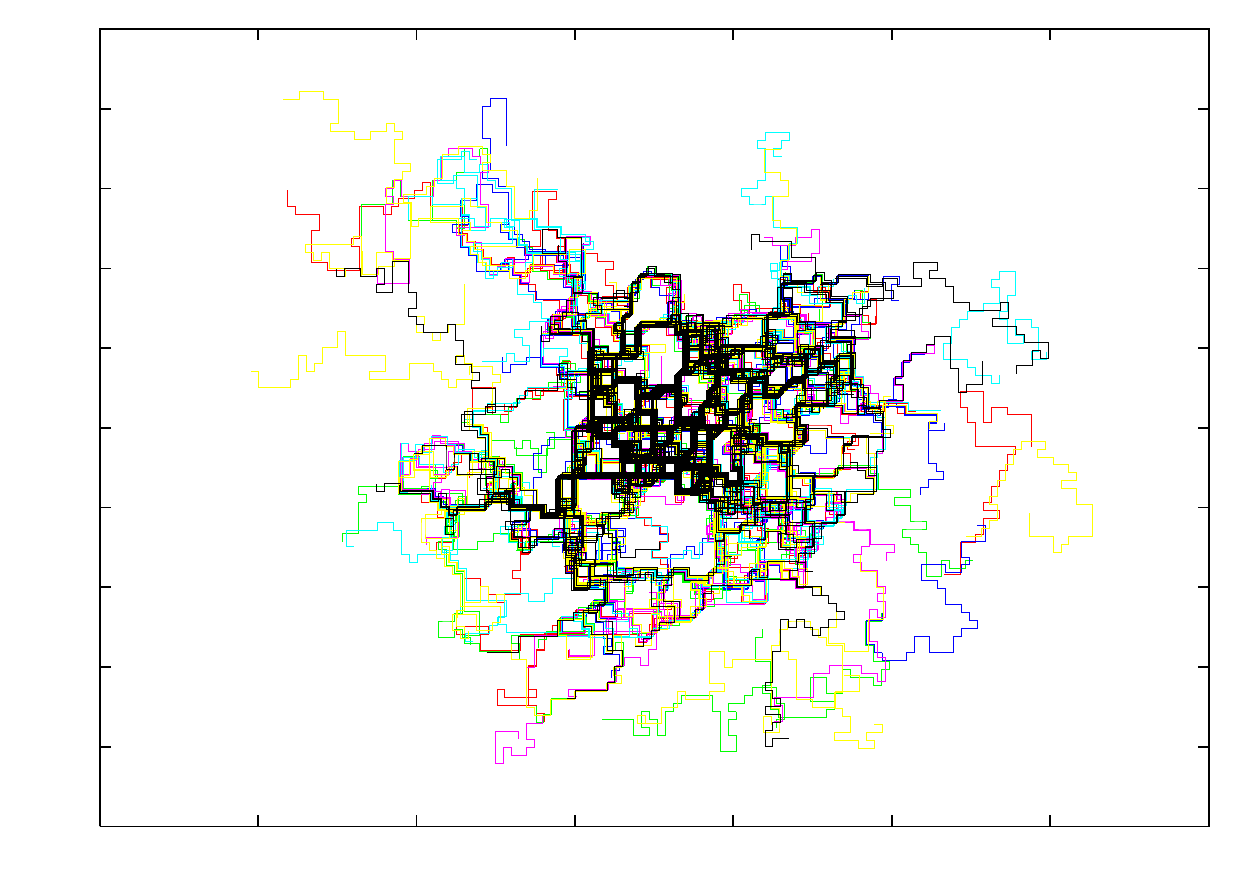} 
	}
	\caption[PTN map of a simulated city]
	{PTN map of a simulated city of with 1024  routes of 64 stations each \cite{Ferber09a}. \label{fig6.2}}
\end{figure}

The above observation gave rise to further exploiting the analogy with SAW and resulted in a model of city PTN. In the framework of the model suggested in 
\cite{Ferber09a} and further exploited in \cite{Holovatch11a} an essential feature of how a PTNs grow is attributed to the attachment of station sequences 
that represent new routes, each of them being modeled as a SAW. 
In this respect the model differs from the standard network grows models, such as preferential attachment type models where networks increase due to 
adding new nodes  \cite{Barabasi99a,Barabasi99b}. Another obvious feature of the model is its embedding in the two-dimensional space. One of the examples 
of simulated PTN maps for a city with 1024 public transport routes of 64 stations each is shown in Fig. \ref{fig6.2}. Such a network growth model captures 
many special features of real world PTNs.

It is instructive to note here, that the `mass' of a RW or of a SAW  expressed as its chemical length $N$, scales with its typical size ${\mathcal R}$ as a power 
law with the non-integer value of the exponent $\nu$, see Eqs. (\ref{6.1}), (\ref{6.2}). In this respect random walks differ from the usual $d$-dimensional objects, 
where the scaling holds with the exponent equal to space dimensionality. Indeed both RW and SAW are well established examples of fractals \cite{Feder88}. The above 
discussed scaling in $d=2$ space brings about their fractal dimensionalities $d_f=2$ and $d_f=4/3$ for RW and SAW correspondingly. The concept of fractals has been 
actively exploited in quantitative description of different communication systems. For the PTNs  and railway networks, fractal structures have been 
analyzed for different subnetworks within Lyon \cite{Thibault87}, Stuttgart    \cite{Frankhauser90}, Paris \cite{Benguigui91,Benguigui92}, several Rhine towns
\cite{Benguigui92}, Seul \cite{Kim03}. These papers analyzed the density of stations, the total length of track as function of the distance from the center of the 
network, the mean distance as a function of number of stations traveled. The distributions of inter-station distances of consecutive PTN stations were shown to have
power-law tails compatible with a L\'evy-flight model \cite{vonFerber13}. 

\begin{figure}[ht!]
	\centerline{
		\includegraphics[width=12cm]{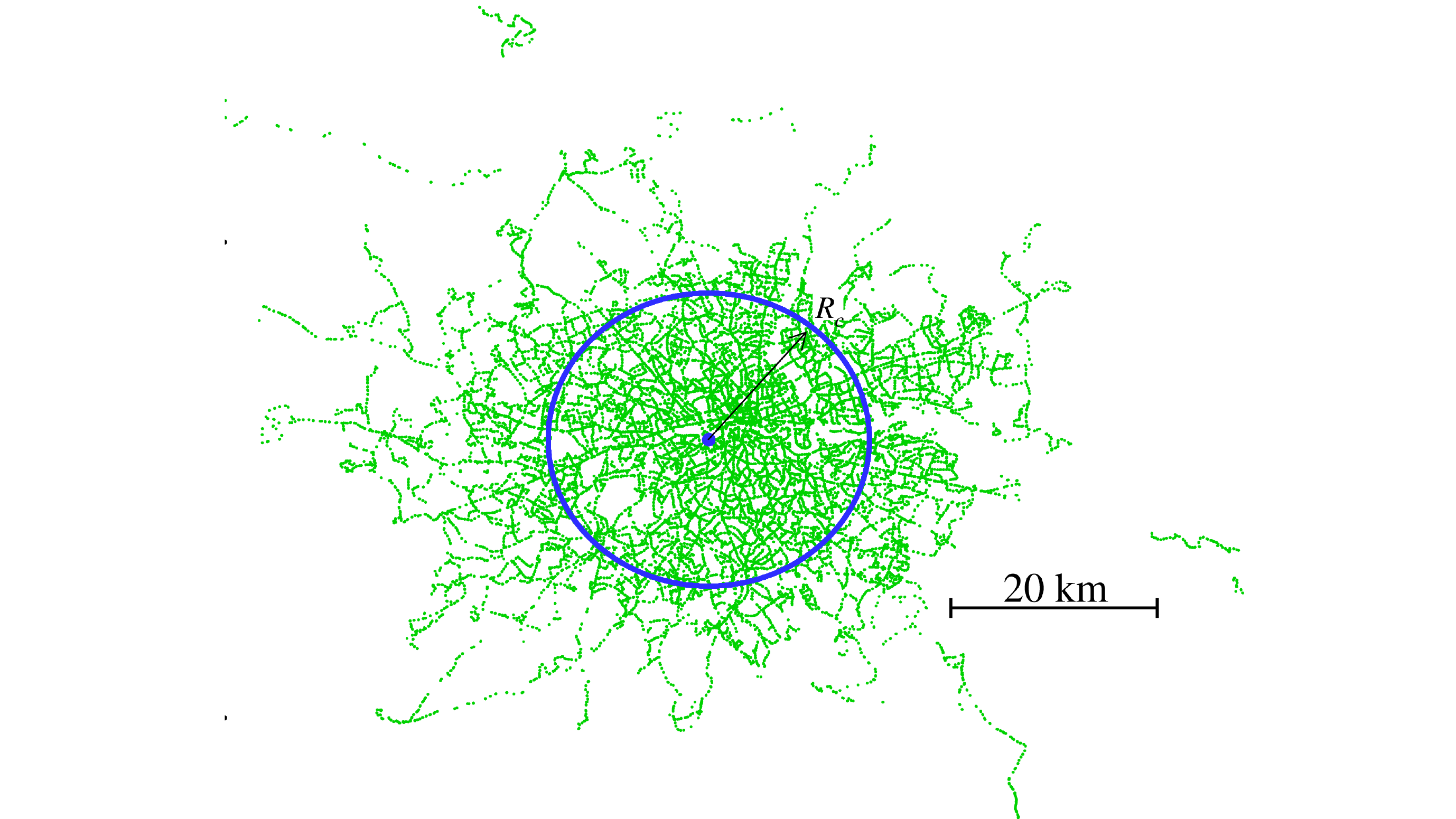} 
	}
	\caption[London PTN]
	{London PTN. The radius $R_c \simeq 15.4$ km corresponds to the transition
from the compact central area to the rarefied space with $d_f < 2$ \cite{deRegt19} . \label{fig6.3}}
\end{figure}

Recently, the fractal dimension of several PTNs in Great Britain has been considered providing a useful interpretation on a PTN serviceability. This opens a 
procedure to apply fractal dimensionality as a key performance indicator and provides additional characteristics of a PTN functional effectiveness \cite{deRegt19}. 
To this end, one investigates the `mass' (number of stations) of the network $N(R)$ as a function of the radius $R$ about PTN center. It was found, that for a small value of $R<R_c$ the PTN tends to cover uniformly the city area while the inhomogeneities in structure are observed at 
the peripheral area. In the central area one observes scaling $N\sim R^2$ whereas $N\sim R^{d_f}$ for $R>R_c$. Such behavior was found to be universal for all PTNs under analysis, whereas the value of $R_c$ was system-dependent, differing for different PTNs. In Fig. \ref{fig6.3} we show an example of such behavior observed for London PTN with $R_c\simeq 15.4$ km \cite{deRegt19}.

\section{Conclusions and outlook}\label{VII}

Physics is an archtype example of a natural science; with its use of mathematics, intricate interplay of experiment-theory-simulation 
and primacy for experiment, well equipped to lend itself to other fields of science. For these reasons it comes as no surprise that over the 
last few decades physics (or at least physicists) has spread into a much wider field of analyzing problems which traditionally might have been 
thought to be the subject of other sciences. Here, physics seeks to better understand how simple rules lead to the diverse and cooperative 
behavior found in complex systems \cite{Holovatch17,Thurner18}. 
Such complex systems can be found in biology, sociology and in other disciplines where interactions between agents play important roles. 
Since all complex systems involve cooperative behavior between many interconnected components, the fields of phase transitions and critical 
phenomena and complex networks give a very natural conceptual and methodological framework for their study. As we aimed to show in this paper, 
utilizing a `physical' approach to study a `unphysically'/'man-made' system of interacting agents provides an effective methodology to contribute 
to a better understanding and prognosis of their collective behavior.

We would like to thank Christian von Ferber,
Bertrand Berche, Taras Holovatch, 
Vasyl Palchykov,  and Mykola Lebovka for common work on studying
public transportation networks as complex systems. 
This work was supported in part by  the project No 0113U003059 of Ministry of Education and Science of Ukraine
(Yu.H.).

\providecommand{\bysame}{\leavevmode\hbox to3em{\hrulefill}\thinspace}
\providecommand{\MR}{\relax\ifhmode\unskip\space\fi MR }
\providecommand{\MRhref}[2]{%
	\href{http://www.ams.org/mathscinet-getitem?mr=#1}{#2}
}
\providecommand{\href}[2]{#2}


\begin{thebibliography}{10}
	
	\bibitem{Albert02}
	R~Albert and A-L Barab\'asi, \emph{Statistical mechanics of complex networks},
	Rev. Mod. Phys. \textbf{74} (2002), 47--97.
	
	\bibitem{alessandretti2016user}
	L Alessandretti, M Karsai, and L Gauvin, \emph{User-based
		representation of time-resolved multimodal public transportation networks},
	Open Science \textbf{3} (2016), \newline 160156.
	
	\bibitem{Anderson72}
	P~W Anderson, \emph{More is different}, Science \textbf{177} (1972), 393--396.
	
	\bibitem{Barabasi99a}
	A-L Barab\'asi and R~Albert, \emph{Emergence of scaling in random networks},
	Science \textbf{286} (1999), 509--512.
	
	\bibitem{Barabasi99b}
	A-L Barab\'asi, R~Albert, and H~Jeong, \emph{Mean-field theory for scale-free
		random networks}, Physica A \textbf{272} (1999), 173--187.
	
	\bibitem{Barthelemy11}
	M Barth\'elemy, \emph{Spatial networks}, Physics Reports \textbf{499}
	(2011), no.~1, 1 -- 101.
	
	\bibitem{Benguigui92}
	L~Benguigui, \emph{The fractal dimension of some railway networks}, Journal de
	Physique \textbf{2} (1992), 385--388.
	
		
	\bibitem{Benguigui91}
	L~Benguigui and M~Daoud, \emph{Is the suburban railway system a fractal?},
	Geographical Analysis \textbf{23} (1991), 362--368.
	
	\bibitem{benguigui1995fractal}
	L Benguigui, \emph{A fractal analysis of the public transportation system
		of Paris}, Environment and Planning A \textbf{27} (1995), 1147--1161.
	
	\bibitem{benguigui1992fractal}
	L Benguigui, \emph{The fractal dimension of some railway networks},
	Journal de Physique I \textbf{2} (1992), 385--388.
	
	\bibitem{berche2009network}
	B Berche, C von Ferber, and T Holovatch, \emph{Network
		harness: bundles of routes in public transport networks}, AIP Conference
	Proceedings, vol. 1198, AIP, 2009, pp.~3--12.
	
	\bibitem{berche2009resilience}
	B Berche, C {von Ferber}, T Holovatch, and Yu~Holovatch,
	\emph{Resilience of public transport networks against attacks}, The European
	Physical Journal B \textbf{71} (2009), no.~1, 125--137.
	
	\bibitem{berche2012transportation}
	B Berche, C {von Ferber}, T Holovatch, and Yu Holovatch,
	\emph{Transportation network stability: a case study of city transit},
	Advances in Complex Systems \textbf{15} (2012), no.~supp01, 1250063.
	
	\bibitem{Bollobas01}
	B Bollob\'as, \emph{Random graphs}, 2 ed., Cambridge Studies in Advanced
	Mathematics, Cambridge University Press, 2001.
	
	\bibitem{chang2007assortativity}
	H Chang, B Su, Y Zhou, and D He, \emph{Assortativity and act
		degree distribution of some collaboration networks}, Physica A: Statistical
	Mechanics and its Applications \textbf{383} (2007), 687--702.
	
	\bibitem{Corominas-Murtra15}
	B~Corominas-Murtra, R~Hanel, and S~Thurner, \emph{Understanding scaling through
		history-dependent processes with collapsing sample space}, Proc. Natl. Acad.
	Sci. USA \textbf{112} (2015), 5348--5353.
	
	\bibitem{Kozitsky14}
	A Daletskii, Yu Kondratiev, Yu Kozitsky, and Tanja Pasurek, \emph{A
		phase transition in a quenched amorphous ferromagnet}, Journal of Statistical
	Physics \textbf{156} (2014), 156--176.
	
	\bibitem{deRegt17}
	R de~Regt, \emph{Complex networks: topology, shape and spatial embedding},
	Doctoral Thesis, Coventry University, Coventry, UK, 2017.
	
	\bibitem{deRegt19}
	R {de Regt}, C {von Ferber}, Yu Holovatch, and M Lebovka,
	\emph{Public transportation in Great Britain viewed as a complex network},
	Transportmetrica A: Transport Science \textbf{15} (2019), no.~2, 722--748.
	
	\bibitem{Dorogovtsev03}
	S~N Dorogovtsev and J~F~F Mendes, \emph{Evolution of networks}, Oxford
	University Press, Oxford, 2003.
	
	\bibitem{Essam80}
	J~W Essam, \emph{Percolation theory}, Reports on Progress in Physics
	\textbf{43} (1980), no.~7, 833--912.
	
	\bibitem{ester1996density}
	M Ester, H-P Kriegel, J Sander, and X Xu, \emph{A
		density-based algorithm for discovering clusters a density-based algorithm
		for discovering clusters in large spatial databases with noise}, Proceedings
	of the Second International Conference on Knowledge Discovery and Data
	Mining, KDD'96, AAAI Press, 1996, pp.~226--231.
	
	\bibitem{Feder88}
	J Feder, \emph{Evolution of networks}, Springer US, 1988.


	\bibitem{vonFerber13}
	C~{von Ferber} and Yu~Holovatch, \emph{Fractal transit networks: self-avoiding
		walks and L\'evy flights}, The European Physical Journal ST \textbf{216}
	(2013), 49--55.
	
	\bibitem{vonFerber12}
	C von Ferber, B Berche, T Holovatch, and Y Holovatch,
	\emph{A tale of two cities}, Journal of Transportation Security \textbf{5}
	(2012), 199--216.
	
	\bibitem{Ferber09a}
	C {von Ferber}, T Holovatch, Yu~Holovatch, and V~Palchykov,
	\emph{Public transport networks: empirical analysis and modeling}, The
	European Physical Journal B \textbf{68} (2009), no.~2, 261--275.
	
	\bibitem{von2007network}
	C {von Ferber}, T Holovatch, Yu~Holovatch, and V Palchykov,
	\emph{Network harness: Metropolis public transport}, Physica A: Statistical
	Mechanics and its Applications \textbf{380} (2007), 585--591.
	
	\bibitem{von2009attack}
	C {von Ferber}, T Holovatch, and Yu Holovatch, \emph{Attack
		vulnerability of public transport networks}, Traffic and Granular Flow’07,
	Springer, 2009, pp.~721--731.
	
	\bibitem{von2009modeling}
	C {von Ferber}, T Holovatch, Yu Holovatch, and V Palchykov,
	\emph{Modeling metropolis public transport}, Traffic and Granular Flow’07,
	Springer, 2009, pp.~709--719.

	
	\bibitem{Frankhauser90}
	P~Frankhauser, \emph{Aspects fractals des structures urbaines}, L'Espace
	G\'eographique \textbf{19} (1990), 45--69.
	
	\bibitem{gallotti2015multilayer}
	R Gallotti and M Barthelemy, \emph{The multilayer temporal network of
		public transport in Great Britain}, Scientific data \textbf{2} (2015),
	140056.
	
	\bibitem{ghosh2010structure}
	S Ghosh, A Banerjee, N Sharma, S Agarwal, A
	Mukherjee, and Niloy Ganguly, \emph{Structure and evolution of the indian
		railway network}, Summer Solstice International Conference on Discrete Models
	of Complex Systems (SOLSTICE), 2010.
	
	\bibitem{Goldenfeld99}
	N~Goldenfeld and L~P Kadanoff, \emph{Simple lessons from complexity.}, Science
	\textbf{284} (1999), 87--89.
	
	\bibitem{guida2007topology}
	M Guida and F Maria, \emph{Topology of the italian airport network:
		A scale-free small-world network with a fractal structure?}, Chaos, Solitons
	\& Fractals \textbf{31} (2007), 527--536.
	
	\bibitem{guimera2004modeling}
	R Guimera and L A~Nunes Amaral, \emph{Modeling the world-wide airport
		network}, The European Physical Journal B-Condensed Matter and Complex
	Systems \textbf{38} (2004), 381--385.
	
	\bibitem{guimera2005worldwide}
	R Guimera, S Mossa, A Turtschi, and L Nunes Amaral, \emph{The
		worldwide air transportation network: Anomalous centrality, community
		structure, and cities' global roles}, Proceedings of the National Academy of
	Sciences of USA \textbf{102} (2005), 7794--7799.
	
	\bibitem{guo2013scaling}
	L Guo, Y Zhu, Z Luo, and W Li, \emph{The scaling of several
		public transport networks in China}, Fractals \textbf{21} (2013), 1350010.
	
	\bibitem{Holovatch11a}
	T Holovatch, \emph{Complex transportation networks: resilience, modelling
		and optimization}, Doctoral Thesis, {Universit{\'e} Henri Poincar{\'e} -
		Nancy 1 and Coventry University}, 2011.
	
	\bibitem{Holovatch06}
	Yu~Holovatch, C~{von Ferber}, A~Olemskoi, T~Holovatch, O~Mryglod, I~Olemskoi,
	and V~Palchykov, \emph{Complex networks}, Journ. Phys. Stud. (in Ukrainian)
	\textbf{10} (2006), 247--291.
	
	\bibitem{Holovatch_books}
	Yu Holovatch (ed.), \emph{Order, disorder and criticality. Advanced problems
		of phase transition theory}, vol. I-V, World Scientific, Singapore,
	2004-2018.
	
	\bibitem{Holovatch17}
	Yu Holovatch, R Kenna, and S Thurner, \emph{Complex systems:
		physics beyond physics.}, European Journal of Physics \textbf{314} (2017),
	023002.
	
	\bibitem{Kadanoff66}
	L P Kadanoff, \emph{Scaling laws for Ising models near $t_c$}, Physics
	Physique Fizika \textbf{2} (1966), 263--272.
	
	\bibitem{Kim03}
	K~S Kim, S~K Kwan, L~Benguigui, and M~Marinov, \emph{The fractal structure of
		Seoul’s public transportation system}, Cities \textbf{20} (2003), 31--39.
	
	\bibitem{Yarynka}
	Ya Korduba, \emph{Public transport networks: Topological features and
		stability analysis}, Bachelor Thesis, Ukrainian Catholic University, Lviv,
	2019.
	
	\bibitem{Kozitsky04}
	Yu Kozitsky, \emph{Mathematical theory of the Ising model and its
		generalizations: an introduction}, Order, Disorder and Criticality. Advanced
	Problems of Phase Transition Theory. Ed. by Yu. Holovatch, World Scientific,
	2004, pp.~1--66.
	
	\bibitem{Kozitsky18}
	Yu Kozitsky and K Pilorz, \emph{Random jumps and coalescence in the
		continuum: evolution of states of an infinite system}, preprint
	arXiv:1807.07310 [math.DS] (2018).
	
	\bibitem{Kozitsky88}
	Yu~V Kozitsky, \emph{Hierarchical model of a vector ferromagnet ---
		self-similar block-spin distributions and the Lee-Yang theorem}, Reports on
	Mathematical Physics \textbf{26} (1988), no.~3, 429 -- 445.
	
	\bibitem{Ladyman13}
	J~Ladyman, J~Lambert, and K~Wiesner, \emph{What is a complex system?}, Eur.
	Journ. Philos. Sci. \textbf{3} (2013), 33--67.
	
	\bibitem{latora2002boston}
	V Latora and M Marchiori, \emph{Is the Boston subway a small-world
		network?}, Physica A: Statistical Mechanics and its Applications \textbf{314}
	(2002), no.~1-4, 109--113.
	
	\bibitem{Mandelbrot53}
	B~Mandelbrot, \emph{An informational theory of the statistical structure of
		languages}, Communication Theory (Woburn, MA) (W~Jackson, ed.), Butterworth,
	1953, pp.~486--502.
	
	\bibitem{Mitzenmacher03}
	M~Mitzenmacher, \emph{A brief history of generative models for power law and
		lognormal distributions}, Internet Mathematics \textbf{1} (2004), 226--251.
	
	\bibitem{molloy1995critical}
	M Molloy and B Reed, \emph{A critical point for random graphs with a
		given degree sequence}, Random structures \& algorithms \textbf{6} (1995),
	no.~2-3, 161--180.
	
	\bibitem{Newman06}
	M~Newman, A-L Barab\'asi, and D~J Watts, \emph{The structure and dynamics of
		networks}, Princeton University Press, Princeton, New Jersey, 2006.
	
	\bibitem{Newman05}
	M~E~J Newman, \emph{Power laws, Pareto distributions and Zipf's law},
	Contemporary Physics \textbf{46} (2005), 323--351.
	
	\bibitem{Nienhuis82}
	B Nienhuis, \emph{Exact critical point and critical exponents of
		$\mathrm{O}(n)$ models in two dimensions}, Phys. Rev. Lett. \textbf{49}
	(1982), 1062--1065.
	
	\bibitem{Parisi99}
	G~Parisi, \emph{Complex systems: a physicist's viewpoint.}, Physica A
	\textbf{263} (1999), 557--564.
	
	\bibitem{Russo17}
	G Russo, V Nicosia, and V Latora, \emph{Complex networks:
		Principles, methods and applications}, Cambridge university press, Cambridge,
	2017.
	
	\bibitem{Schneider11}
	C~M Schneider, A~A Moreira, J~S Andrade, S Havlin,
	and Hans~J Herrmann, \emph{Mitigation of malicious attacks on networks},
	Proceedings of the National Academy of Sciences \textbf{108} (2011), no.~10,
	3838--3841.
	
	\bibitem{Seaton2004}
	K Seaton and L Hackett, \emph{Stations, trains and small-world
		networks}, Physica A \textbf{339} (2004), 635--644.
	
	\bibitem{sen2003small}
	P Sen, S Dasgupta, A Chatterjee, P~Sreeram, G~Mukherjee, and
	S~Manna, \emph{Small-world properties of the Indian railway network},
	Physical Review E \textbf{67} (2003), no.~3, 036106.
	
	\bibitem{Sengupta06}
	A Sengupta (ed.), \emph{Chaos, nonlinearity, complexity: The dynamical
		paradigm of nature}, Sringer, 2006.
	
	\bibitem{Shanmukhappa16}
	T Shanmukhappa, I W H Ho, and C~K Tse, \emph{Bus transport
		network in hong kong: Scale-free or not?}, 2016 International Symposium on
	Nonlinear Theory and Its Applications, NOLTA2016, Yugawara, Japan, November
	27th-30th, 2016, pp.~610--614.
	
	\bibitem{Sherrington10}
	D~Sherrington, \emph{Physics and complexity.}, Phil. Trans. Roy. Soc. A
	\textbf{368} (2010), 1175--1189.
	
	\bibitem{Sienkiewicz05}
	J Sienkiewicz and J Ho{\l}yst, \emph{Statistical analysis of 22
		public transport networks in Poland}, Physical Review E \textbf{72} (2005),
	46127.
	
	\bibitem{Simkin11}
	M~V Simkin and V~P Roychowdhury, \emph{Re-inventing Willis}, Physics Reports
	\textbf{502} (2011), 1--35.
	
	\bibitem{Simon55}
	H~A Simon, \emph{On a class of skew distribution functions}, Biometrika
	\textbf{42} (1955), 425--440.
	
	\bibitem{soh2010weighted}
	H Soh, S Lim, T Zhang, X Fu, G Lee,
	T Hung, P Di, S Prakasam, and L Wong,
	\emph{Weighted complex network analysis of travel routes on the singapore
		public transportation system}, Physica A: Statistical Mechanics and its
	Applications \textbf{389} (2010), 5852--5863.
	
	\bibitem{Stanley71}
	H~E Stanley, \emph{Introduction to phase transitions and critical phenomena},
	Oxford University Press, 1971.
	
	\bibitem{Stauffer91}
	D~Stauffer and A~Aharony, \emph{Introduction to percolation theory}, Taylor \&
	Francis, London, 1991.
	
	\bibitem{sui2012space}
	Y~Sui, F Shao, R Sun, and S Li, \emph{Space evolution
		model and empirical analysis of an urban public transport network}, Physica
	A: Statistical Mechanics and its Applications \textbf{391} (2012),
	3708--3717.
	
	\bibitem{Thibault87}
	S~Thibault and A~Marchand (eds.), \emph{R\'eseaux et topologie}, Institut
	National des Sciences Applique\'es de Lyon, Villeurbanne, 1987.
	
	\bibitem{Thurner18}
	S~Thurner, R~Hanel, and P~Klimek, \emph{Introduction to the theory of
		complex systems}, Oxford University Press, 2018.
	
	\bibitem{Thurner16}
	S Thurner (ed.), \emph{Visions for complexity}, World Scientific,
	Singapore, 2016.

	\bibitem{xu2007scaling}
	X Xu, J Hu, F Liu, and L Liu, \emph{Scaling and
		correlations in three bus-transport networks of China}, Physica A:
	Statistical Mechanics and its Applications \textbf{374} (2007), 441--448.
	
	\bibitem{yang2011bus}
	X H Yang, G Chen, B Sun, S Y Chen, and W L Wang, \emph{Bus
		transport network model with ideal n-depth clique network topology}, Physica
	A: Statistical Mechanics and its Applications \textbf{390} (2011),
	4660--4672.
	
\end{thebibliography}
\end{document}